\documentclass{article}

\usepackage[page,header]{appendix}
\usepackage[dvipsnames]{xcolor}
\usepackage{natbib, amssymb, amsmath, amsthm, times, graphicx, tikz, comment}
\usepackage{thmtools}
\usepackage{algorithm}
\usepackage[noend]{algpseudocode}
\usepackage{booktabs, caption, makecell}
\usepackage{subfig}
\usepackage{authblk}
\usepackage{scrextend}
\usepackage{hyperref}
\usepackage{enumitem}
\usepackage{threeparttable}
\usepackage[margin=3.5cm]{geometry}
\usepackage[parfill]{parskip}
\usetikzlibrary{datavisualization, plotmarks, arrows.meta, shadings}

\newcommand{\To}{\textbf{ to }}

\title{eTFCE: Exact Threshold-Free Cluster Enhancement via Fast Cluster Retrieval}
\author[1]{Xu Chen \thanks{Wivenhoe Park, Colchester CO4 3SQ, UK; xc23776@essex.ac.uk.}}
\author[2]{Wouter D. Weeda}
\author[3,4]{Thomas E. Nichols}
\author[5]{Jelle J. Goeman}
\affil[1]{School of Mathematics, Statistics and Actuarial Science, University of Essex, UK}
\affil[2]{Methodology and Statistics Unit, Department of Psychology, Leiden University, The Netherlands}
\affil[3]{Big Data Institute, Li Ka Shing Centre for Health Information and Discovery, Nuffield Department of Population Health, University of Oxford, UK}
\affil[4]{Wellcome Centre for Integrative Neuroimaging, FMRIB, Nuffield Department of Clinical Neurosciences, University of Oxford, UK}
\affil[5]{Biomedical Data Sciences, Leiden University Medical Center, The Netherlands}
\date{\today}

\begin{document}

\maketitle

\begin{abstract}
\noindent Threshold-free cluster enhancement (TFCE) is widely used for cluster-based inference in neuroimaging, but existing implementations typically rely on discretized approximations that may introduce numerical variability. We present eTFCE, an efficient framework that provides a numerically exact evaluation of the TFCE integral using an optimized cluster retrieval algorithm. Across multiple datasets, eTFCE and the standard implementation produce highly consistent inference results. Voxel-wise comparisons reveal a systematic asymmetry: the standard method yields smaller $p$-values for more voxels, while eTFCE concentrates stronger statistical evidence within a smaller subset. These differences are primarily confined to voxels near the inference boundary and have minimal impact on overall inference. This pattern is consistent with discretization effects in standard implementations, where the TFCE integral is approximated using a finite set of threshold levels, introducing subtle biases in statistical evidence accumulation across thresholds. Furthermore, eTFCE improves computational efficiency ($71.3\%$ of runtime on average) and enables unified computation of multiple cluster-based statistics within a single permutation framework. Overall, eTFCE provides an exact, efficient, and extensible approach to nonparametric neuroimaging inference.
\end{abstract}

\section{Introduction}

In neuroimaging, traditional cluster extent inference suffers from issues of dependence on arbitrarily chosen cluster defining thresholds (CDTs) and low spatial specificity \citep{Woo2014,Eklund2016,Rosenblatt2018}. Threshold-free cluster enhancement (TFCE) mitigates the impact of these limitations by integrating spatial clustering information and voxel-wise intensity across all possible threshold levels \citep{Smith2009,Salimi-Khorshidi2011}. However, TFCE is computationally intensive, as it theoretically requires both the exact evaluation of the integral-based TFCE score over all threshold levels and repeated computations within a nonparametric permutation testing framework \citep{Nichols2002,Winkler2014}. In practice, most implementations rely on discrete approximations of the TFCE integral, which may deviate from the exact solution, while differences in thresholding schemes can introduce additional variability across results.

This paper addresses the challenges of computational efficiency and numerical accuracy by introducing an exact and efficient TFCE implementation (eTFCE). In standard practice, the TFCE integral is discretized over a fixed set of CDT levels (e.g.,\ $100$ equally spaced CDTs, as used by default in FSL \citep{Jenkinson2012}) to reduce computational cost. However, this discretization compromises mathematical exactness and reveals inherent limitations of fixed-step approximations, motivating the development of exact and efficient alternatives.

In practice, a range of efficient TFCE implementations have been proposed to improve computational performance in large-scale neuroimaging settings, particularly for connectivity-based analyses. These approaches introduce algorithmic and structural optimizations to reduce computational burden while maintaining consistency with the TFCE framework (e.g.,\ IC-TFCE \citep{Cravo2026} and its implementations in the CONN toolbox \citep{Whitfield-Gabrieli2012}). Related computational tools, such as PRISME \citep{Cravo2026JAN}, further provide a general framework for large-scale benchmarking and power analysis in neuroimaging.

Our key insight is that the TFCE integral can be computed more efficiently than previously assumed using an optimized cluster retrieval algorithm \citep{Chen2023}, removing dependence on discretized approximations. The proposed eTFCE produces mathematically exact TFCE scores while yielding results that remain consistent with standard implementations, and requires approximately $71.3\%$ of the computation time of the commonly used FSL implementation. Thus, it improves the computational efficiency without altering the underlying statistical properties of TFCE, providing a more efficient  realization of the same inference procedure, which makes eTFCE a practical tool for reliable nonparametric inference in neuroimaging.

Furthermore, eTFCE supports efficient computation of multiple cluster-based statistics through shared intermediate cluster representations within a single permutation framework. In particular, traditional cluster statistics such as cluster extent and cluster mass can be obtained at minimal additional cost along with the TFCE scores. This enables both voxel-wise and cluster-level inference to be derived from a single analysis run. More broadly, this highlights the flexibility of the proposed computational framework, which can accommodate a range of cluster-based inference methods with similar computational structures and constraints, with minimal modification.

\section{Background}

In this section, we briefly review two key components: the TFCE score and the cluster retrieval algorithm.

\subsection{Threshold-Free Cluster Enhancement (TFCE)} \label{sec:tfce}

The TFCE score at voxel $v$ in the analysis brain mask $\mathcal{B}$ is defined as 
\begin{equation}
\text{TFCE}(v) = \int_{h_0}^{h_v} e_v(h)^E h^H \,dh,
\label{intTFCE}
\end{equation}
where $h$ denotes the chosen CDT level ranging from $h_0$ (typically set to $0$) to the voxel height $h_v$. Here, $e_v(h)$ is the size of the cluster, thresholded at level $h$, that contains voxel $v$, and $E$ and $H$ are the extent and height parameters, respectively. \citet{Smith2009} recommended $E=0.5$ and $H=2$ for 3D data, as this choice ensures sensitivity across a wide range of signal levels.

While the TFCE score is defined by the integral \eqref{intTFCE}, common implementations, including the default in FSL, approximate it through discretization over a sequence of non-decreasing CDTs $\{\tau_i\}_{i=1}^n$. Here, the number of thresholds $n$ determines the step size of the approximation. This yields a general discrete approximation of the TFCE score:
\begin{equation}
\text{TFCE}(v) \approx \sum_{i=1}^n e_v(\tau_i)^E \tau_i^H \Delta \tau_i,
\label{discTFCE}
\end{equation}
where $\Delta \tau_i$ represents the local step size associated with $\tau_i$. Typically, equally spaced thresholds ($\Delta \tau_i = \Delta \tau$) are used with $n$ set to a relatively small number for computational efficiency, which can introduce a relatively large approximation bias.

The default threshold selection strategy implemented in FSL's \texttt{randomise} and \texttt{fslmaths} \citep[version 6.0.7.19; ][]{Jenkinson2012}, as well as in FSL \texttt{PALM}'s TFCE implementation, uses $n=100$ uniformly spaced thresholds between $0$ and $h_{\max}$. Given a predefined number of thresholds $n$, a constant step size is automatically determined as $\Delta \tau = h_{\max} / n$, where $h_{\max}$ denotes the maximum voxel height in $\mathcal{B}$. The TFCE scores are then computed independently at each voxel, with cluster sizes evaluated separately at each threshold $\tau_i$. This discretized approximation may therefore introduce numerical differences in the resulting nonparametric inference.

\subsection{Cluster Retrieval Algorithm} \label{sec:union-find}

To efficiently identify all supra-threshold clusters in statistic or $p$-value maps, we implemented a cluster retrieval algorithm based on the disjoint-set (union-find) data structure. Let $h_v$ denote the statistic at voxel $v$ in the original voxel index space, and let $h_{(i)}$ denote the $i^\text{th}$ value after sorting the statistics in non-ascending order. Similarly, let $p_v$ denote the $p$-value at voxel $v$, and $p_{(i)}$ the $i^\text{th}$ $p$-value after sorting in non-descending order. The cluster retrieval algorithm operates on the sorted sequences $h_{(i)}$ (or $p_{(i)}$) to identify contiguous supra-threshold clusters. 

The procedure mainly follows the method introduced by \citet{Chen2023}. It processes all voxels in a single pass after sorting, merging adjacent supra-threshold voxels into spatially connected clusters through union operations. By doing so, the algorithm can identify and list all supra-threshold clusters across all thresholds in almost linear time.

As outlined in Algorithm \ref{alg} and illustrated in Figure \ref{fig:eTFCE}, the algorithm iteratively constructs a directed rooted forest from an input graph. The input is represented as an undirected graph, where nodes correspond to supra-threshold voxels and edges reflect spatial adjacency under a predefined connectivity criterion (e.g.,\ 4- or 8-connectivity in 2D space, and 6‑, 18‑ or 26‑connectivity in 3D space). 

The algorithm initializes an edgeless forest in which each node is its own root, forming a forest of trivial trees. The original statistics $h_v$ (or $p_v$) are then processed in non-ascending (or non-descending) order. For each node $i$ in the sorted sequence, the algorithm examines all neighboring voxels with smaller rank indices. If such a neighbor belongs to a different subtree, node $i$ points to the root of that subtree, and the two subtrees are merged into a larger one. This process is repeated until all voxels have been visited, yielding a forest that naturally partitions the voxels into connected components (or clusters). 

During the step-by-step forest construction procedure, each rooted subtree corresponds uniquely to a supra‑threshold cluster. The root node serves as the cluster identifier, and all nodes within the subtree define the spatial extent of the cluster. Crucially, the iterative process outlined in Algorithm \ref{alg} directly builds the entire cluster structure, relying on this one-to-one mapping between subtree roots and clusters. As a result, no additional post‑processing is required for cluster identification.

As shown in Figure \ref{fig:eTFCE}, $9$ nodes are sorted in non-ascending order of their statistic values. To illustrate the process, consider the step at which node $(7)$ is processed. At this step, node $(7)$ merges with the existing subtrees rooted at nodes $(5)$ and $(6)$ through its neighbors $(2)$, $(5)$, and $(6)$, resulting in a new subtree rooted at node $(7)$ that constitutes a single cluster. This example illustrates the incremental cluster formation during the forest construction process using Algorithm \ref{alg}.

The algorithm preserves the near-linear time complexity of the original union‑find design through path compression, making it efficient for large-scale neuroimaging data. In our implementation, we use standard neighborhood definitions consistent with existing implementations \citep{Chen2023}, allowing our method to accommodate different spatial assumptions.

\begin{algorithm}[t]
\caption{Construction of a directed rooted forest from an undirected graph (adapted from Algorithm A in \citep{Chen2023}).}
\label{alg}
\begin{algorithmic}[0]
\Require{$h_{(1)} \geq h_{(2)} \geq \cdots \geq h_{(N)}$ or $p_{(1)} \leq p_{(2)} \leq \cdots \leq p_{(N)}$, where $\mathcal{B} = (\mathbb{V}, \mathbb{E})$ is the input graph with node set $\mathbb{V}$ and edge set $\mathbb{E}$, and $N = |\mathbb{V}|$ denotes the number of nodes in $\mathcal{B}$.}
\Statex
\Function{FOREST}{$\mathbb{V}$, $\mathbb{E}$}
	\State Initialize an edgeless directed rooted forest $\mathcal F$ on $\mathbb{V}$.
    \Statex
	\For{$i = 1$ \To $N$}
		\ForAll{$\{j,i\} \in \mathbb{E}$ such that $j < i$}
				\State Find the root $r$ of the subtree in $\mathcal F$ that contains $j$.
				\If{$i \neq r$}
			 		\State Add edge $(i, r)$ to $\mathcal F$.
				\EndIf
		\EndFor
	\EndFor
	\Statex
	\State \Return{$\mathcal{F}$}  
\EndFunction
\end{algorithmic}
\end{algorithm}

\begin{figure}[htbp]
\centering
\begin{tikzpicture}[
    graphnode/.style={circle,draw, inner sep=1pt, minimum size=16pt, align=center},
    labelnode/.style={below}, scale=1.2]
    \node[graphnode](A) at (-2.5,3.6) {$12.5$\\\small(1)}; 
    \node[graphnode](B) at (-0.7,3.6) {$4.1$\\\small(5)}; 
    \node[graphnode](C) at (1.1,3.6) {$7.3$\\\small(4)}; 
    \node[graphnode](D) at (-2.5,1.8) {$2.1$\\\small(8)}; 
    \node[graphnode](E) at (-0.7,1.8) {$2.9$\\\small(7)}; 
    \node[graphnode](F) at (1.1,1.8) {$10.2$\\\small(2)}; 
    \node[graphnode](G) at (-2.5,0) {$9.8$\\\small(3)}; 
    \node[graphnode](H) at (-0.7,0) {$3.5$\\\small(6)}; 
    \node[graphnode](I) at (1.1,0) {$1.2$\\\small(9)}; 
    
    \path (A) edge (B) edge (D);
    \path (E) edge (B) edge (D) edge (F) edge (H);
    \path (G) edge (D) edge (H);
    \path (I) edge (H) edge (F);
    \path (C) edge (B) edge (F);
  
    \begin{scope}[xshift=4cm, yscale=0.6, xscale=0.9, >={Stealth[round]}]
        \node[graphnode, align=center](1) at (0,0) {$12.5$\\\small(1)}; 
        \node[graphnode, align=center](2) at (3,0) {$10.2$\\\small(2)}; 
        \node[graphnode, align=center](3) at (5,0) {$9.8$\\\small(3)};
        \node[graphnode, align=center](4) at (2,1) {$7.3$\\\small(4)};
        \node[graphnode, align=center](5) at (1,2) {$4.1$\\\small(5)};
        \node[graphnode, align=center](6) at (4,2) {$3.5$\\\small(6)}; 
        \node[graphnode, align=center](7) at (2.5,4) {$2.9$\\\small(7)};
        \node[graphnode, align=center](8) at (3.75,5) {$2.1$\\\small(8)}; 
        \node[graphnode, align=center](9) at (5,6) {$1.2$\\\small(9)};  
        
        \path[->] (9) edge (8); 
        \path[->] (8) edge (7);
        \path[->] (7) edge (5) edge (6);        
        \path[->] (5) edge (1) edge (4);
        \path[->] (6) edge (3);
        \path[->] (4) edge (2);        
    \end{scope}
    \node[labelnode] at (-0.8,-0.8) {(a) Original graph};
    \node[labelnode] at (6.5,-0.8) {(b) Final Forest};
\end{tikzpicture}
\caption{Illustration of Algorithm \ref{alg} for constructing a directed rooted forest. (a) Input undirected graph with $9$ nodes, where each node label shows the statistic value (top) and the corresponding sorting index (bottom), indicating the processing order. (b) Output directed rooted forest after applying the 2D edge-connectivity (or $4$-connectivity) criterion.}
\label{fig:eTFCE}
\end{figure}

\section{eTFCE: Exact TFCE with Efficient Cluster Retrieval}

In this section, we first detail the novel integration of exact TFCE formulation with cluster retrieval algorithm that accelerates computation while preserving accuracy, and then present a generalized formula for cluster statistics including TFCE.

\subsection{Exact TFCE Formulation}

We first simplify the computation of the TFCE integral \eqref{intTFCE}. Traditional implementations approximate this integral through numerical discretization with a small number of uniform thresholds. In contrast, our approach utilizes the piecewise constant nature of $e_v(h)$ to evaluate the integral analytically. 

Specifically, $e_v(h)$ is zero for $h > h_v$, and changes only at certain voxel heights $h \le h_v$. Let all in-mask voxels with positive height be assigned rank indices $1,2,\ldots,N$ according to their heights in non-increasing order, where $N$ is the number of such voxels in $\mathcal{B}$. For each voxel $v$, we define an index set $\Phi_v \subseteq \{1,2,\ldots,N\}$, whose elements correspond to rank indices in this sorted sequence. The smallest element in $\Phi_v$ corresponds to $v$ itself, while the remaining indices indicate the ranks at which the size of the cluster containing $v$ changes. This induces a non-decreasing sequence of CDT levels $\{\tau_i\}_{i=1}^{\mid \Phi_v \mid}$, where each $\tau_i$ is given by the height of the voxel with the corresponding rank index in $\Phi_v$, and $|\cdot|$ denotes the cardinality of a set. In particular, $\tau_{\mid \Phi_v \mid}=h_v$. By further defining $\tau_0=0$, we obtain the full CDT sequence $\{\tau_i\}_{i=0}^{\mid \Phi_v \mid}$.

With each interval $\big( \tau_{i-1}, \tau_i \big]$, $1 \le i \le | \Phi_v |$, the cluster extent $e_v(h)$ remains constant and equals $e_v(\tau_i)$. Because $e_v(h)$ changes only at thresholds $\tau_i$, the TFCE integral can be decomposed over these intervals and we obtain an exact discrete representation:
\begin{equation}
\text{TFCE}(v) = \sum_{i=1}^{\mid \Phi_v \mid} \int_{\tau_{i-1}}^{\tau_i} e_v(h)^E h^H \,dh = \frac{1}{H+1} \sum_{i=1}^{\mid \Phi_v \mid} e_v(\tau_i)^E \Big(\tau_i^{H+1} - \tau_{i-1}^{H+1}\Big).
\label{eTFCE}
\end{equation}
This formulation eliminates the numerical errors introduced by discrete approximation. Its exactness follows directly from the piecewise constant property of $e_v(h)$, which remains constant on every interval $\big(\tau_{i-1}, \tau_i\big]$ for $1 \le i \le | \Phi_v |$.

The main challenge in applying Equation \eqref{eTFCE} lies in efficiently determining $\Phi_v$ for each voxel $v$, which would be computationally costly if performed naively. However, by using Algorithm \ref{alg} described in Section \ref{sec:union-find} and the directed rooted forest it constructs, this process becomes highly efficient. Consequently, our approach removes the step size dependence present in conventional approximations (e.g.,\ the fixed default of $100$ thresholds in FSL) while maintaining computational efficiency.

\subsection{Efficient Cluster Retrieval Integration}

We now integrate the exact TFCE formula \eqref{eTFCE} with the efficient cluster retrieval algorithm (Algorithm \ref{alg}) to accelerate the exact TFCE computation (see Algorithm \ref{alg2}). This integration is built on a disjoint-set data structure, which detects all supra-threshold clusters in near-linear time with respect to the number of voxels \citep{Chen2023}.

In the eTFCE framework, a directed rooted forest that encodes the hierarchical voxel relationships is first constructed using a disjoint-set data structure (see the $9$-node example in Figure \ref{fig:eTFCE}). During the union operations in Algorithm \ref{alg}, key cluster properties, such as cluster extent and root voxel, are recorded. Building upon this pre-constructed forest, Algorithm \ref{alg2} computes the TFCE score for each voxel intrinsically during cluster identification, i.e.,\ while determining its index set $\Phi_v$. Specifically, $\Phi_v$ is obtained by tracing the path from the node (or voxel $v$) to its root in the forest. E.g.,\ when processing node $(4)$ in Figure \ref{fig:eTFCE}, $\Phi_4 = \{4,5,7,8,9\}$ is obtained by following the path from node $(4)$ to its root $(9)$. Each node along this path corresponds to a CDT level at which the cluster extent of node $(4)$ changes. This procedure generalizes to all voxels, and enables simultaneous cluster retrieval and TFCE computation across all effective thresholds in a single pass, thus eliminating the need for repeated evaluations at individual thresholds.

By utilizing the pre-constructed forest, this integrated design achieves both computational efficiency and mathematical exactness. The disjoint-set data structure accelerates the most computationally intensive phase of TFCE computation, while automatically capturing all cluster transitions at unique voxel heights without explicit thresholding. In contrast to standard approaches that rely on approximate discretization for enhanced efficiency or decreasing step sizes for improved accuracy, our proposed method evaluates the TFCE integral exactly over all data-adaptive thresholds, thus avoiding approximation bias.

\begin{algorithm}[t]
\caption{Exact TFCE calculation for all voxels using the directed rooted forest from Algorithm \ref{alg}.}
\label{alg2}
\begin{algorithmic}[0]
\Require{$h_{(1)} \geq h_{(2)} \geq \cdots \geq h_{(N)}$, where $\mathcal{F}$ is the pre-constructed directed rooted forest with $N$ nodes, implemented as a disjoint-set data structure.}
\Statex
\Function{TFCE}{$\mathcal{F}$}
	\State Initialize a TFCE map $T$ as all zeros.
    \Statex
	\For{$i = N$ \To $1$}
        \State $u = \mathrm{parent}_\mathcal{F}(i)$ \Comment{Parent of node $i$ in forest $\mathcal{F}$}
        \If{$u \neq i$} \Comment{Check if a root node is encountered}
            \State $\Delta t_i = \frac{1}{H+1} e_i\big(h_{(i)}\big)^E \Big(h_{(i)}^{H+1} - h_{(u)}^{H+1}\Big)$
            \State $T(i) = T(u) + \Delta t_i$ \Comment{Update based on Equation \eqref{eTFCE}}
        \Else
            \State $T(i) = \frac{1}{H+1} e_i\big(h_{(i)}\big)^E h_{(i)}^{H+1}$
        \EndIf
	\EndFor
	\Statex
	\State \Return{$T$}
\EndFunction
\end{algorithmic}
\end{algorithm}

\subsection{Nonparametric Inference Procedure}

The underlying computational framework of eTFCE also incorporates nonparametric inference through resampling (permutation or sign flipping). The choice of appropriate randomization strategy depends on the experimental design, e.g.,\ sign flipping is typically used for one-sample tests, while permutation (or label shuffling) is used for tests involving two independent samples \citep{Nichols2002,Andreella2023}.

The corresponding nonparametric resampling procedure implemented in eTFCE is detailed below.
\begin{enumerate}
\item \textbf{Compute original TFCE map.} \\ 
Apply the exact TFCE formulation, together with the cluster retrieval algorithm, to the original data to obtain the observed TFCE map $T_\mathrm{orig}$.
\item \textbf{Generate null distribution via resampling.} \\
For each resampling iteration $b=1,\ldots,N_\mathrm{perm}$:
\begin{itemize}
\item Randomize the data according to the experimental design (e.g.,\ randomly permuting group labels, or applying random sign flips).
\item Compute the corresponding TFCE map $T^{(b)}$ using the exact TFCE formulation combined with the cluster retrieval algorithm.
\item Record the global maximum:
\begin{equation*}
t_{\max}^{(b)} = {\max}_{v \in \mathcal{B}} \ T^{(b)}(v), 
\end{equation*}
where $T^{(b)}(v)$ denotes the TFCE score at voxel $v$ for randomization $b$.
\end{itemize}
\item \textbf{Construct empirical null distribution.} \\
The empirical null distribution is formed by the set of maximum TFCE scores $\big\{ t_{\max}^{(b)} \big\}_{b=1}^{N_\mathrm{perm}}$.
\item \textbf{Calculate TFCE $p$-values.} \\
For each voxel $v \in \mathcal{B}$, compute
\begin{equation*}
p_v^\mathrm{FWE} = \frac{\#\big\{ b \colon t_{\max}^{(b)} \ge T_\mathrm{orig}(v) \big\} + 1}{N_\mathrm{perm} + 1}.
\end{equation*}
Voxels can then be thresholded at a desired $\alpha$-level (e.g.,\ $p_v^\mathrm{FWE} \le 0.05$).
\end{enumerate}

This eTFCE framework offers enhanced reliability and practicality of statistical inference. By preserving the analytical exactness of the TFCE integral across all randomizations, it eliminates variability introduced by discrete approximations. Additionally, each resampling iteration employs data-adaptive height thresholds, ensuring that the empirical null distribution is fully determined by the intrinsic data structure without requiring any externally specified step sizes. Together with the almost linear time complexity of the cluster retrieval algorithm, this approach enables large-scale nonparametric inference within a consistent and computationally efficient framework.

In practice, existing TFCE implementations differ in how the integral is discretized. Standard approaches rely on discrete approximations, but the discretization strategy is not always consistent across software tools. For instance, FSL's \texttt{randomise} uses a fixed step size determined from the first permutation and applies it uniformly across all permutations, while FSL's \texttt{PALM} fixes the number of steps and normalizes the resulting statistic to ensure comparability across multiple data modalities.

These differences imply that the TFCE integral does not have a unique discrete implementation. Consequently, the TFCE-based inference may exhibit subtle variability depending on the chosen discretization scheme. In contrast, the proposed eTFCE framework avoids such implementation-dependent effects by directly evaluating the TFCE integral in an exact, data-adaptive manner.

\subsection{Generalized Formulation for Cluster Statistics} \label{sec:genTFCE}

Beyond the specific task of accelerating TFCE computation, we note that TFCE belongs to a broader family of cluster statistics that integrate information from both voxel height and spatial cluster extent. These methods can be expressed within a generalized formulation:
\begin{equation*}
T(v) = \int_{h=h_0}^{\infty} g\big(e_v(h)\big) f(h) \,dh,
\end{equation*}
where $f(\cdot)$ and $g(\cdot)$ are functions of the height threshold $h$ and the corresponding cluster extent $e_v(h)$ at voxel $v$, respectively. 

Under this formulation, standard TFCE is defined by the specific choices: $f(x) = x^H$ and $g(x) = x^E$. Notably, this framework is highly adaptable. By appropriately specifying both functions, a range of classical cluster-based inference methods can be reproduced, including voxel-wise peak height inference, cluster-extent inference, and cluster-mass inference. Specifically, they can be achieved using:
\begin{itemize}
\item Voxel-wise peak height inference: $f(x) = 1$ and $g(x) = \mathbf{1}_{\{x>0\}}$, where $\mathbf{1}_{\{x>0\}}$ is an indicator function for $x>0$.
\item Cluster-extent inference: $f(x) = \delta(x-h_v)$ and $g(x) = x$, where $\delta(x-h_v)$ is the Dirac delta function that is equal to $0$ except at $x=h_v$.
\item Cluster-mass inference: $f(x) = 1$ and $g(x) = x$.
\end{itemize}

Our proposed eTFCE approach, combining exact TFCE evaluation with an efficient cluster retrieval algorithm, naturally extends to this broader family of cluster-level statistics. As a result, the framework can alleviate the computational burden associated with a wide range of inference methods, including probabilistic TFCE (pTFCE) \citep{Spisak2019} and localized cluster enhancement (LCE) \citep{Davenport2024,Weeda2025}.

\section{Empirical Evaluation}

We now compare the proposed eTFCE approach with the most commonly used TFCE implementation in FSL's \texttt{randomise} as the baseline, which employs a fixed uniform threshold sequence (see details in Section \ref{sec:tfce}).

\subsection{Experimental Setup}

We evaluate our method against FSL's TFCE implementation in \texttt{randomise} (version 6.0.7.19) in terms of statistical performance and computational efficiency. Here, statistical performance is assessed by the similarity of the resulting TFCE-corrected $p$-values, while computational efficiency is measured as the average per-randomization runtime for all voxels. Evaluations are conducted on an fMRI auditory dataset \citep{Pernet2015} and six task contrasts from the Human Connectome Project \citep[HCP; ][]{Barch2013,VanEssen2013} to assess both methodological validity and generalizability.

To ensure a fair comparison, sign-flipping permutations generated by \texttt{randomise} are reused in the eTFCE pipeline, ensuring identical randomization schemes across methods. For each of the $5000$ randomizations, a single sign-flipping pattern is applied consistently at the subject level and shared across all voxels, preserving the within-subject spatial dependence structure. Using these shared permutations, both methods compute group-level one-sample $t$-statistic maps, followed by TFCE transformation. Whole-brain null distributions are then constructed from the resulting TFCE scores, from which TFCE-corrected $p$-values are derived. The statistical validity of applying sign flipping to contrast maps has been well-established \citep{Andreella2023}.

While both pipelines include additional steps such as image loading and data handling, these are present in both implementations and therefore contribute comparably to the overall runtime. As a result, the measured per-randomization times reflect realistic end-to-end performance under consistent conditions. This setup ensures that any observed differences in computational efficiency arise from the TFCE-based nonparametric implementations rather than differences in randomization or input data, which enables a controlled and practically meaningful comparison.

\subsection{Data}

We describe the datasets used to evaluate the performance of eTFCE using two types of datasets to assess both methodological validity and generalizability.

First, we used a publicly available auditory dataset \citep{Pernet2015}, in which $140$ healthy participants performed a vocal vs non-vocal sound discrimination task. Individual functional images were preprocessed using SPM12 following the pipeline described in \citet{Pernet2015}. First-level contrast maps (vocal $>$ non-vocal) were computed for each subject and analyzed at the group level using a one-sample $t$-test to obtain a whole-brain $t$-statistic map.

To further examine performance across different activation patterns \citep{Noble2020}, we additionally analyzed six cognitive task contrasts from the Human Connectome Project \citep[HCP; ][]{Barch2013,VanEssen2013}, following the evaluation framework of \citet{Noble2020}. These contrasts include:
\begin{itemize}
\item Emotional processing: faces vs shapes,
\item Incentive processing (gambling): punish vs reward,
\item Language processing (story): math vs story,
\item Relational processing: matching vs relational,
\item Social cognition: theory of mind (ToM) vs random,
\item N-back Working memory: 2-back vs 1-back.
\end{itemize}
Data from $80$ independent subjects were extracted from the HCP S1200 data release, and preprocessed using the minimal preprocessing pipelines described by \citet{Glasser2013}. First-level contrast maps were computed and entered into group-level analysis to generate a one-sample $t$-statistic map.

\subsection{Statistical Performance}

\begin{figure}[htbp]
\centering
\includegraphics[width=\textwidth]{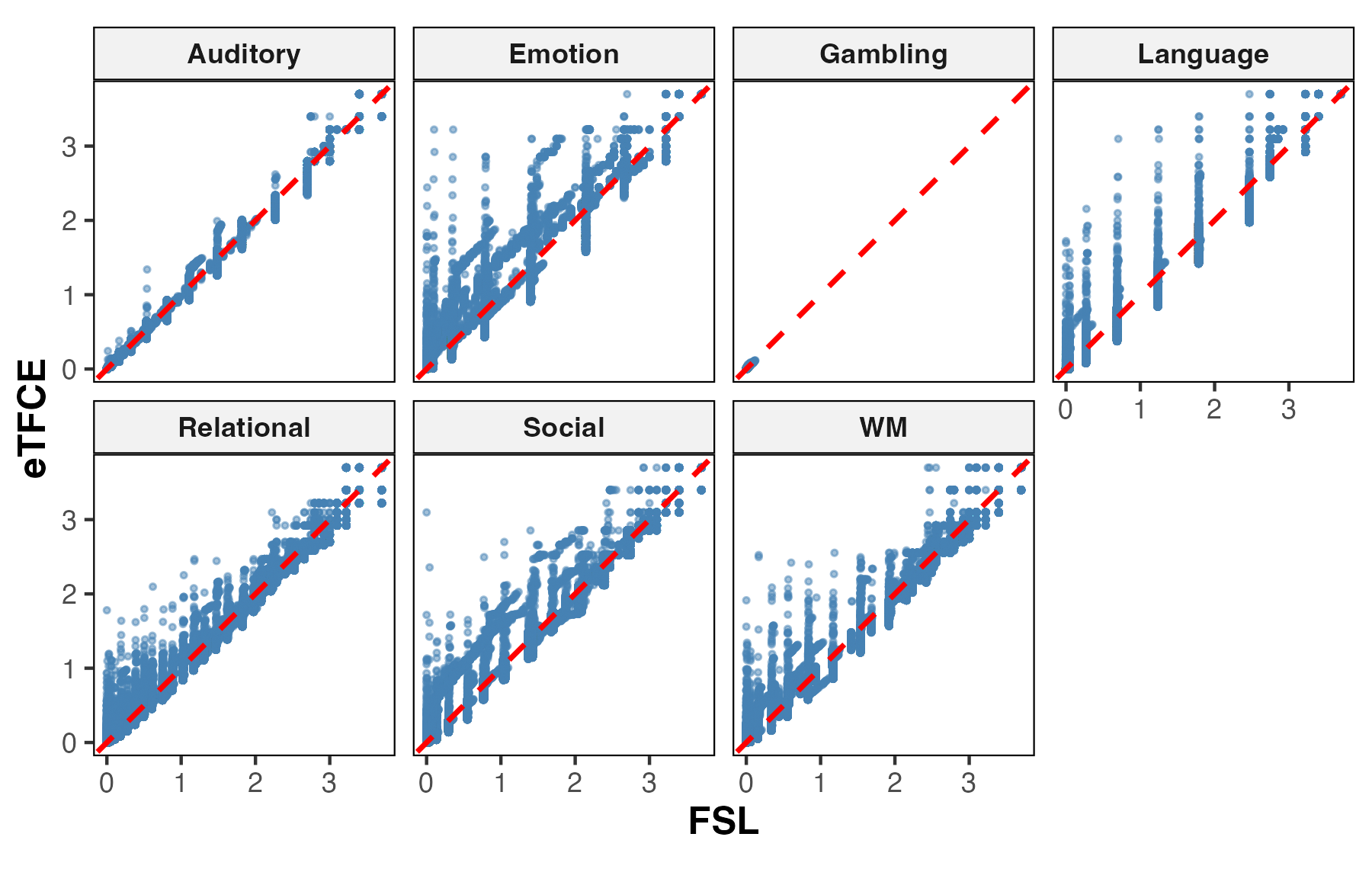}
\caption{Comparison of eTFCE and FSL's default TFCE on the auditory task data (vocal vs non-vocal) and $6$ cognitive task contrasts from the Human Connectome Project (HCP). The log-transformed TFCE-corrected $p$-values ($-\log_{10}(p)$, based on $5000$ sign-flipping randomizations) are shown. The red dashed lines indicate the identity line, and the six HCP contrasts are: Emotional (faces vs shapes), Gambling (punish vs reward), Language (math vs story), Relational (matching vs relational), Social (ToM vs random), and WM (2-back vs 1-back).} 
\label{fig:performance} 
\end{figure}

Figure \ref{fig:performance} summarizes the comparison between eTFCE and FSL's default TFCE implementation across all datasets by showing voxel-wise relationships in log-transformed TFCE-corrected $p$-values. Overall, both methods exhibit strong agreement, with the majority of voxels lying close to the identity line. For the Auditory dataset, deviations are small and approximately symmetric, while for the Gambling contrast, both methods yield uniformly large $p$-values, resulting in effectively identical inference.

Table \ref{tbl:performance} further quantifies voxel-wise and inference-level differences across all task contrasts. Although voxel-wise discrepancies in $p$-values are present, a consistent asymmetry is observed. While FSL yields smaller $p$-values for a larger proportion of voxels, the magnitude of these differences is generally modest. In contrast, when eTFCE produces more significant results, the corresponding differences in $-\log_{10}(p)$ are consistently larger (Mean $\text{D}^+ >$ Mean $|\text{D}^-|$ across all contrasts). This indicates that eTFCE tends to produce stronger effects in a smaller subset of voxels, while FSL more frequently yields marginal increases in significance.

Importantly, these differences do not translate into substantial changes in binary inference outcomes. Gain and loss rates remain low across all contrasts, typically below $1\%$, with a maximum loss of approximately $1.5\%$ observed in the Relational task. Notably, these discrepancies are predominantly localized near the inference boundary, rather than within strongly significant clusters, suggesting that they reflect reclassification of marginal voxels rather than meaningful changes in detected activation patterns.

For the remaining HCP task contrasts (e.g.,\ Emotion, Language, Relational, Social, and Working Memory), the voxel-wise distributions reveal a consistent asymmetric pattern. Rather than reflecting a global shift between methods, this asymmetry corresponds to a localized redistribution of statistical evidence. Specifically, eTFCE selectively amplifies a subset of voxels with intermediate-to-strong signal, while FSL exhibits a broader but weaker pattern of voxel-wise significance.

This behavior can be attributed to differences in the numerical implementation of TFCE. The standard TFCE implementations are based on discretized threshold integration, which introduces quantization effects and limits the resolution of TFCE values. As a result, voxel-wise statistics exhibit reduced dispersion and increased clustering near threshold boundaries, leading to a higher proportion of voxels close to significance thresholds with marginally smaller $p$-values. In contrast, eTFCE evaluates the TFCE integral exactly, preserving the continuous contribution of cluster extent across CDT levels. This allows stronger signals, particularly those associated with high-threshold regions and complex cluster geometries, to be more accurately represented, resulting in larger effect sizes for a subset of voxels.

Consistent with this interpretation, in Table \ref{tbl:performance}, larger differences between methods are observed in task contrasts characterized by more complex spatial organization (e.g.,\ Emotion and Relational), where cluster extent changes rapidly across CDT levels. In such cases, discretization effects are enlarged, leading to increased voxel-wise discrepancies. Conversely, in contrasts with simpler or weaker activation patterns (e.g.,\ Gambling), both methods produce nearly identical results.

Overall, eTFCE and FSL yield highly consistent TFCE-corrected inference results across all datasets. The primary difference lies in how statistical evidence is distributed voxel-wise. FSL tends to produce more widespread but weaker marginal effects, while eTFCE yields more concentrated and stronger effects in a smaller subset of voxels. Crucially, these differences are largely restricted in boundary regions and do not alter the overall pattern of statistical inference.

\begin{table}[htbp]
\centering
\begin{threeparttable}
\caption{Voxel-wise comparison of TFCE-based TFCE-corrected $p$-values between eTFCE and FSL, obtained using $5000$ sign-flipping permutations. $\text{D}$ is defined as the voxel-wise difference between $-\log_{10}(p_\text{eTFCE})$ and $-\log_{10}(p_\text{FSL})$. $\text{D}^+$ and $\text{D}^-$ denote the positive and negative voxel-wise differences, respectively. Gain and Loss quantify the proportions of in-mask voxels that change TFCE significance status between methods (significant under eTFCE but not FSL, and vice versa).}
\begin{tabular}{@{\extracolsep{\fill}} l r r r r r r r}
\toprule
Contrast & $\text{D}^+$ ($\%$) & $\text{D}^-$ ($\%$) & Mean $\text{D}^+$ & Mean $|\text{D}^-|$ & Gain ($\%$) & Loss ($\%$) \\
\midrule
Auditory       &  $7.8$ & $21.8$ & $0.051$ & $0.028$ & $0.1$ & $0.1$ \\
Emotion        & $11.0$ & $24.0$ & $0.120$ & $0.061$ & $0.3$ & $1.0$ \\
Gambling       &  $0.8$ &  $0.6$ & $0.002$ & $0.001$ & $0.0$ & $0.0$ \\
Language       &  $8.3$ & $49.2$ & $0.094$ & $0.038$ & $0.5$ & $0.0$ \\
Relational     &  $6.8$ & $34.9$ & $0.089$ & $0.074$ & $0.1$ & $1.5$ \\
Social         &  $5.6$ & $23.0$ & $0.149$ & $0.035$ & $0.2$ & $0.3$ \\
Working Memory &  $4.6$ & $30.9$ & $0.105$ & $0.050$ & $0.1$ & $0.3$ \\
\bottomrule
\end{tabular}
\begin{tablenotes}
\footnotesize
\item Note: $\text{D} = \log_{10}(p_\text{FSL}) - \log_{10}(p_\text{eTFCE})$. $\text{D}^+$ and $\text{D}^-$ ($\%$) denote the proportion of in-mask voxels where $\text{D} > 0$ and $\text{D} < 0$, respectively. Mean $\text{D}^+$ and Mean $|\text{D}^-|$ denote the mean positive and mean absolute negative voxel-wise differences.
\end{tablenotes}
\label{tbl:performance}
\end{threeparttable}
\end{table}

\subsection{Computational Efficiency}

In addition to the statistical performance results presented above, we evaluate the computational efficiency of eTFCE and FSL's default TFCE implementation on the auditory task and six HCP task contrasts (Table \ref{tbl:time}). We measure the running time required to compute the whole-brain TFCE map from each randomization, and report the average running time per randomization over $5000$ sign-flipping permutations.

The reported runtimes for both methods include voxel-wise statistic computation and the corresponding inference procedure. For eTFCE, the total runtime further includes disjoint-set forest construction, TFCE evaluation, and cluster-level inference (cluster extent and cluster mass) within a unified computational framework, together with sorting of the permutation-specific statistic map. In contrast, FSL's \texttt{randomise} reports overall runtimes for TFCE, cluster extent (CE), and cluster mass (CM) procedures separately. Due to its black-box implementation, a finer breakdown of internal computational components is not available.

Across all task contrasts, eTFCE consistently achieves lower overall runtime than FSL's TFCE implementation. On average, eTFCE requires approximately $71.3\%$ of the runtime of FSL. This difference is observed despite eTFCE performing additional computations within each permutation, including forest construction and exact TFCE computation, while FSL employs a discretized approximation of the TFCE statistic.

A key factor contributing to the computational efficiency of eTFCE is its integrated design. The disjoint-set forest representation enables efficient topological information aggregation and supports the simultaneous computation of TFCE and cluster-level statistics within a unified computational framework. Although forest construction is performed separately for each permutation due to the permutation-specific ordering of statistics, the underlying connectivity structure is pre-computed once and reused, allowing efficient reconstruction of the disjoint-set representation within each iteration.

Overall, these results demonstrate that eTFCE provides an efficient and unified implementation of TFCE-based inference while maintaining exact computation of the TFCE integral and enabling joint access to cluster-level statistics within a single analysis pipeline.

\begin{table}[htbp]
\centering
\begin{threeparttable}
\caption{Average running time (seconds) per randomization for whole-brain analysis over $5000$ permutations. Reported runtimes for both eTFCE and FSL include voxel-wise statistic computation and the corresponding inference procedure. eTFCE integrates TFCE and cluster-level inference within a unified framework, while FSL reports runtimes for TFCE, cluster extent (CE), and cluster mass (CM) separately.}
\begin{tabular}{@{\extracolsep{\fill}} l r r r r r r r r}
\toprule
& & \multicolumn{4}{c}{eTFCE} & \multicolumn{3}{c}{FSL} \\
\cmidrule(lr){3-6} \cmidrule(lr){7-9}
Contrast & Voxels$^{1}$ & Total$^{2}$ & Forest$^{3}$ & TFCE & CE\& CM$^{4,5}$ & TFCE & CE$^4$ & CM$^4$ \\
\midrule
Auditory       & $168211$ & $0.310$ & $0.078$ & $0.001$ & $0.000$ & $0.454$ & $0.368$ & $0.373$ \\
Emotion        & $257659$ & $0.326$ & $0.130$ & $0.002$ & $0.000$ & $0.402$ & $0.329$ & $0.336$ \\
Gambling       & $257659$ & $0.325$ & $0.130$ & $0.002$ & $0.000$ & $0.586$ & $0.327$ & $0.334$ \\
Language       & $257659$ & $0.324$ & $0.130$ & $0.002$ & $0.000$ & $0.396$ & $0.331$ & $0.337$ \\
Relational     & $257659$ & $0.324$ & $0.130$ & $0.002$ & $0.000$ & $0.486$ & $0.329$ & $0.334$ \\
Social         & $257659$ & $0.326$ & $0.130$ & $0.002$ & $0.000$ & $0.423$ & $0.327$ & $0.335$ \\
Working Memory & $257659$ & $0.325$ & $0.130$ & $0.002$ & $0.000$ & $0.424$ & $0.328$ & $0.331$ \\
\bottomrule
\end{tabular}
\begin{tablenotes}
\footnotesize
\item $^1$The voxel count within the mask is reported.
\item $^2$eTFCE total runtime includes voxel-wise statistic computation, disjoint-set forest construction, TFCE evaluation, and cluster-level inference. 
\item $^3$Forest construction also includes sorting of the statistic map.
\item $^4$Supra-threshold clusters are defined using a CDT of $h \ge 3.1$. CE and CM denote cluster extent and cluster mass, respectively.
\item $^5$The reported time includes the combined computation of both cluster extent and cluster mass.
\end{tablenotes}
\label{tbl:time}
\end{threeparttable}
\end{table}

\subsection{Integrated TFCE and Cluster-Level Inference}

\begin{figure}[htbp]
\centering
\subfloat[CMI-defined significant clusters (blue shading) overlaid on TFCE-significant voxels (hot colormap).]{\label{fig:visual-cmi}\includegraphics[width=\textwidth]{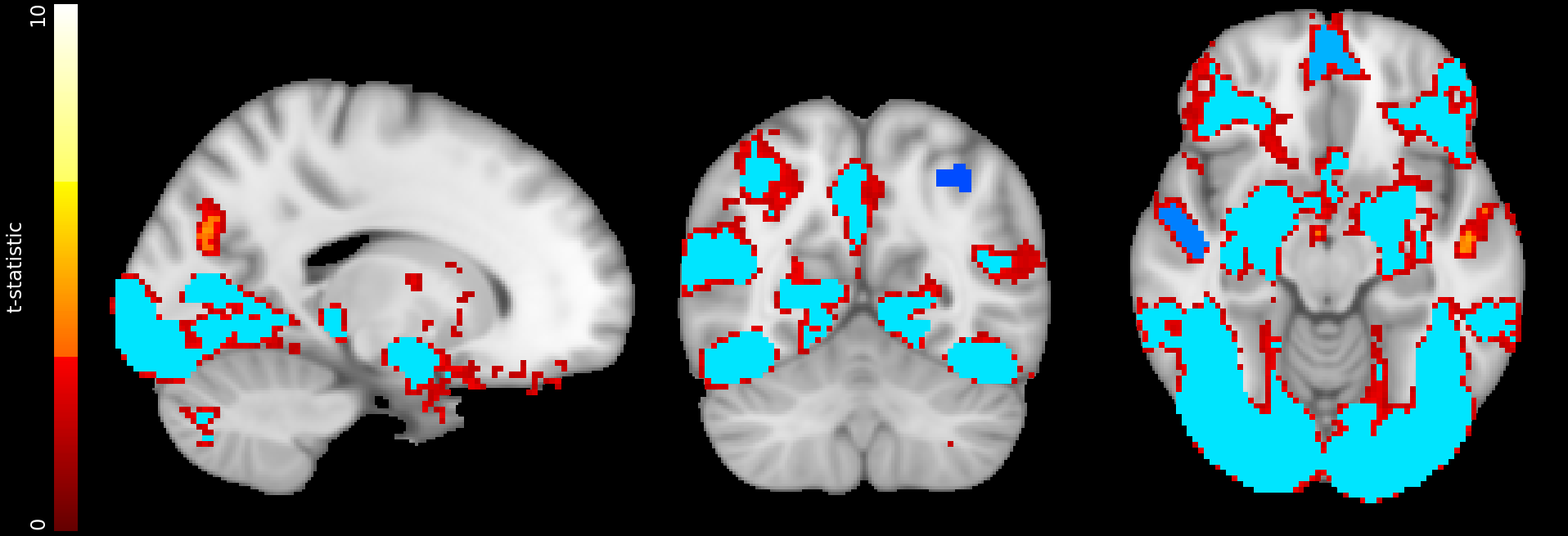}}
\par\medskip
\centering
\subfloat[TFCE-significant voxels (hot colormap) overlaid on CMI-defined clusters (blue shading).]{\label{fig:visual-tfce}\includegraphics[width=\textwidth]{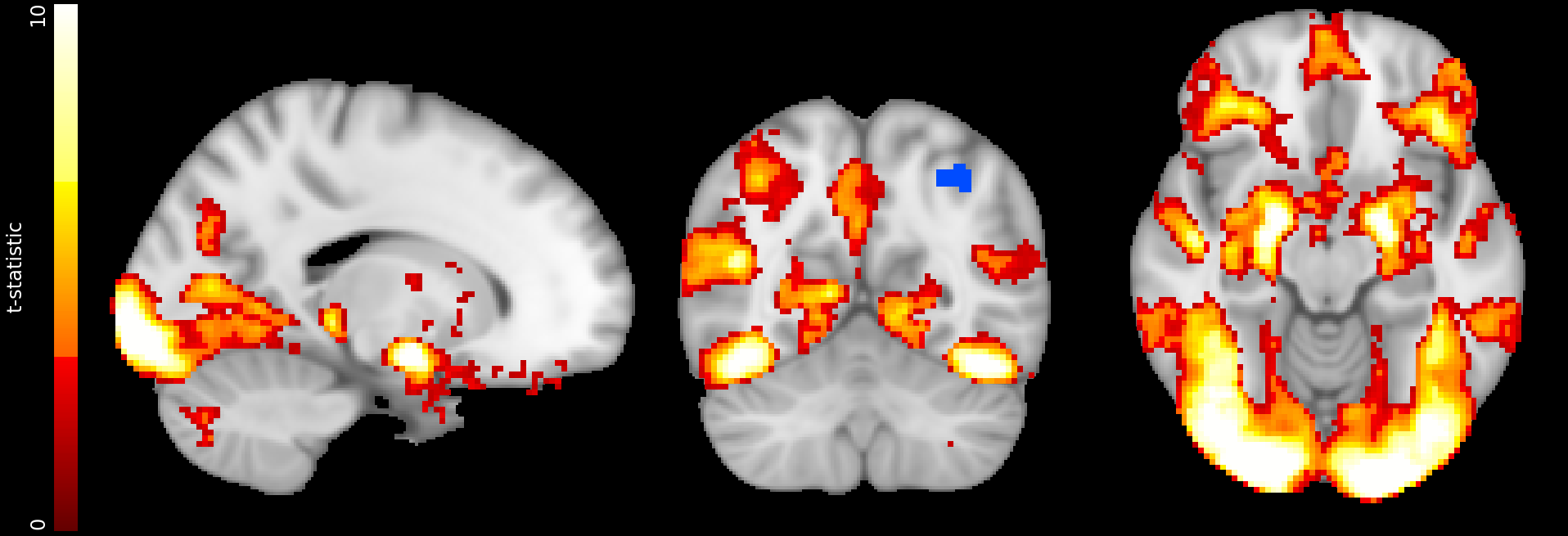}}
\caption{Example visualization of integrated TFCE and cluster mass inference (CMI) on HCP Emotion contrast (faces $>$ shapes) using eTFCE. The figure illustrates the spatial correspondence between voxel-wise TFCE-based inference (TFCE-corrected voxel-wise $p \le 0.05$, hot colormap) and cluster-level CMI results (TFCE-corrected cluster-level $p \le 0.05$, blue shading). The $t$-map is displayed with thresholding determined by TFCE-derived corrected $p$-values.}
\label{fig:visual}
\end{figure}

To characterize spatial patterns of task-evoked activation, we jointly examine voxel-wise TFCE-based inference and cluster-level inference on the HCP Emotion contrast (faces $>$ shapes), as illustrated in Figure \ref{fig:visual}. Both methods are applied independently to the same statistical maps and provide complementary insights into spatial signal organization rather than serving as alternative inferential procedures.

Figure \ref{fig:visual} illustrates a high degree of spatial correspondence between the two approaches across most brain regions, providing a stable baseline for identifying regional differences. We focus on cluster mass inference (CMI) for comparison, as it provides a complementary characterization of spatial structure, while cluster extent inference yields a more conservative estimate of cluster significance, resulting in one fewer detected cluster.

Both methods show broadly consistent effects in the left hemisphere, with activations localized primarily in the occipito-temporal cortex. In contrast, localized differences emerge in the right hemisphere: TFCE shows slightly greater spatial extension into the right fusiform cortex, while CMI identifies a localized cluster in the lateral occipital cortex that is not detected by TFCE.

These region-specific differences are consistently observed across both visualization orientations (see Figure \ref{fig:visual-cmi} \& \ref{fig:visual-tfce}), suggesting that they are not dependent on representation ordering. Rather than indicating inconsistency between methods, these differences reflect their distinct statistical formulations and sensitivity profiles.

Overall, this joint analysis highlights the complementary aspects of TFCE and cluster-level inference in characterizing task-related activation, with TFCE tending to emphasize spatially extended effects and cluster-level inference more strongly reflecting spatially localized clusters.

\section{Conclusion and Discussion}

We propose eTFCE, an exact reformulation of TFCE combined with an efficient cluster retrieval algorithm. Across multiple datasets, eTFCE and the standard FSL implementation show highly consistent voxel-wise inference, with most values concentrated near the identity line, indicating strong overall agreement between the two approaches.

Despite this overall consistency, a systematic voxel-wise asymmetry is observed. In regions where differences occur, FSL tends to yield slightly smaller $p$-values across a larger proportion of voxels, while eTFCE produces stronger effects in a smaller subset of voxels ($\text{D}^+ > |\text{D}^-|$). More precisely, FSL more frequently introduces marginal increases in significance, while eTFCE concentrates statistical strength more locally. Importantly, these differences are spatially limited and primarily occur near inference boundaries, resulting in minimal changes in TFCE-corrected decisions. Consistently, low gain and loss rates across contrasts further indicate that these voxel-wise discrepancies do not translate into meaningful changes under TFCE control, suggesting that eTFCE does not alter inference globally but instead redistributes statistical evidence in a non-uniform, signal-dependent manner.

These differences can be attributed to the underlying numerical formulation. Discretized TFCE employs a fixed discretization of the threshold space, introducing quantization effects that compress the dynamic range of voxel-wise statistics and under-represent contributions from high-threshold regions. In contrast, eTFCE computes TFCE scores via an exact continuous integral formulation, avoiding discretization error and preserving contributions across all thresholds. This results in a more precise representation of strong, spatially localized signals, particularly in regions with complex cluster geometry.

Beyond statistical performance, eTFCE also improves computational efficiency. It requires on average $71.3\%$ of the runtime of the standard implementation, while avoiding discretization approximations in the TFCE integral evaluation. Furthermore, its unified framework enables multiple inference strategies (currently including TFCE, cluster extent, and cluster mass) to be computed within a single nonparametric run with minimal additional cost. This framework can be directly extended to other TFCE variants and related cluster-based statistics with similar computational structure.

Overall, eTFCE provides an exact, efficient, and extensible framework for nonparametric inference in neuroimaging. It preserves voxel-wise consistency with existing implementations while reducing discretization-related approximation bias, lowering computational cost, and providing a unified framework for cluster-based inference.

\section{Acknowledgments}

This work was supported by the London Mathematical Society Scheme 4: Research in Pairs grant (Ref. 42438 to X.C.).

\newpage
\bibliography{refs}
\bibliographystyle{apalike}	

\end{document}